\documentclass[11pt]{article}

\usepackage{microtype}
\usepackage{graphicx}
\usepackage{subcaption}
\usepackage{booktabs}
\usepackage{microsoft-tech-report}
\usepackage{hyperref}
\usepackage{url}
\usepackage[utf8]{inputenc}
\usepackage[T1]{fontenc}
\usepackage{amsmath}
\usepackage{amssymb}
\usepackage{amsfonts}
\usepackage{mathtools}
\usepackage{amsthm}
\usepackage{bm}
\usepackage{nicefrac}
\usepackage{enumitem}
\usepackage{multirow}
\usepackage[capitalize,noabbrev]{cleveref}
\usepackage{xspace}
\usepackage{pifont}
\usepackage{wrapfig}
\usepackage{algorithm}
\usepackage{algpseudocode}
\usepackage{placeins}
\usepackage{lineno}
\usepackage[most]{tcolorbox}
\usepackage[table]{xcolor}

\lstdefinestyle{patchbox}{%
  basicstyle=\scriptsize\ttfamily,
  breaklines=true,
  breakatwhitespace=false,
  breakautoindent=false,
  breakindent=0pt,
  columns=fullflexible,
  keepspaces=true,
  showstringspaces=false,
  postbreak=\mbox{\textcolor{gray}{$\hookrightarrow$}\space},
  aboveskip=0pt,
  belowskip=0pt,
}

\newcommand{\osbench}{\textsc{OS-Bench}\xspace}

\definecolor{deltapos}{HTML}{0E8A4A}
\definecolor{deltaneg}{HTML}{B91C1C}

\techreportlabel{Microsoft Research}
\techreportshorttitle{OS-Bench}

\hypersetup{
  colorlinks=true,
  linkcolor=msftblue,
  citecolor=msftblue,
  urlcolor=msftblue,
  pdftitle={OS-Bench: Evaluating SWE Agents on Cross-OS Code Migration}
}

\begin{document}
\thispagestyle{empty}

\noindent
\begin{minipage}[c]{0.5\linewidth}
\raggedright
\raisebox{-0.5\height}{\msftbrandmark}
\end{minipage}%
\begin{minipage}[c]{0.49\linewidth}
\raggedleft
{\msftdatefont\small\color{msftgray}October 2026}
\end{minipage}\par
\vspace{0.35em}
\noindent{\color{msftline}\rule{\linewidth}{0.8pt}\par}

\vspace{1.0em}
\begin{center}
{{\msfttitlefont\fontsize{21}{25}\selectfont\color{msftdark}
\osbench{}: Evaluating SWE Agents on Cross-OS Code Migration\par}}
\vspace{1.25em}

{\normalsize\rmfamily\color{msftdark}
Qirui Jin$^{1,*,\ddagger}$ \hspace{0.6em}
Kenan Li$^{2,*,\ddagger}$ \hspace{0.6em}
Liao Zhu$^{2}$ \hspace{0.6em}
Lingching Tung$^{2}$ \hspace{0.6em}
Yikai Zhang$^{2}$ \hspace{0.6em} \\
Yijia Wu$^{2}$ \hspace{0.6em} 
Zijian Jin$^{2}$ \hspace{0.6em}
Pengfei Gao$^{2}$ \hspace{0.6em}
Yu Kang$^{2}$ \hspace{0.6em}
Qingwei Lin$^{2}$ \hspace{0.6em}
Wenke Lee$^{1}$\par
}
\vspace{0.22cm}

{\footnotesize\rmfamily\color{msftgray}
$^{1}$ Georgia Institute of Technology \quad
$^{2}$ Microsoft\par
}
\end{center}

\vspace{0.45em}
\begin{msfttitlebox}
\setlength{\parindent}{0cm}
\setlength{\parskip}{0.14cm}
\raggedright
\nohyphens

\begin{abstract}
Language-model (LM) agents are increasingly evaluated on repository-level software engineering (SWE) benchmarks derived from GitHub issues and pull requests.
Although these benchmarks cover a broad range of development tasks, they provide limited insight into a common yet technically distinct challenge: migrating an existing implementation from one operating system (OS) to another while preserving behavior on both.
We introduce \osbench{}, an executable benchmark for cross-OS code migration between Linux and Windows.
Motivated by case studies showing that cross-platform features are frequently implemented on one OS before support is added for another, \osbench{} identifies migration opportunities by comparing execution coverage collected from the same repository on both systems.
Platform-specific regions are then verified, organized into five migration scenarios, and transformed into repository-level tasks with target-platform tests and source-platform regression tests.
The resulting benchmark contains 140 validated tasks from 111 repositories across C, C++, C\#, Go, and Rust, comprising 91 Windows-to-Linux and 49 Linux-to-Windows migrations.
We evaluate six models with SWE-agent, Claude Code, OpenHands, and Win-agent.
The best configurations achieve only 26.4\% success on Linux-target tasks and 20.4\% on Windows-target tasks while preserving correctness across both systems.
These results indicate that cross-OS migration is a challenging and underexplored capability, and \osbench{} is a testbed for improving platform-aware SWE agents.
\end{abstract}

\vspace{0.14cm}
{\setlength{\parskip}{0.06cm}\small
{\msftmetalabel{Dataset}\href{https://huggingface.co/datasets/SWE-bench-Live/OS-bench}{SWE-bench-Live/OS-bench}\par}
{\msftmetalabel{Correspondence}\href{mailto:qiruijin@gatech.edu}{qiruijin@gatech.edu}, \href{mailto:kenan.li.24@ucl.ac.uk}{kenan.li.24@ucl.ac.uk}\par}
}
\vspace{0.08cm}
{\footnotesize\rmfamily\itshape\color{msftgray}
$^{*}$ Equal contribution. \quad $^{\ddagger}$ Corresponding authors.\par
}
\end{msfttitlebox}


\section{Introduction}
Cross-platform support is essential for software libraries, developer tools, and applications that need to run across different computing environments.
Supporting a new operating system (OS), however, requires more than recompiling the same source code.
Developers often need to handle differences in system APIs, filesystem and path conventions, process and signal behavior, networking interfaces, text encodings, build toolchains, and command-line environments.
A migration patch must therefore implement the required behavior on the target platform while preserving existing functionality on the source platform.

LM-based software engineering agents can explore repositories, execute tests, and modify multiple files in response to natural-language instructions.
Their progress is typically evaluated on benchmarks built from real GitHub issues or pull requests~\citep{jimenez2024swe,jin2026chainswe,zhang2026sweedit,li2026oracle}.
While these benchmarks capture realistic repository-level development tasks, they usually evaluate each task in only one execution environment.
As a result, they do not directly test whether an agent can identify an OS-specific implementation gap, adapt behavior across incompatible platform abstractions, work effectively in the target operating system, and preserve existing functionality on the original OS.

\begin{figure*}[t]
\centering
\includegraphics[width=0.95\linewidth]{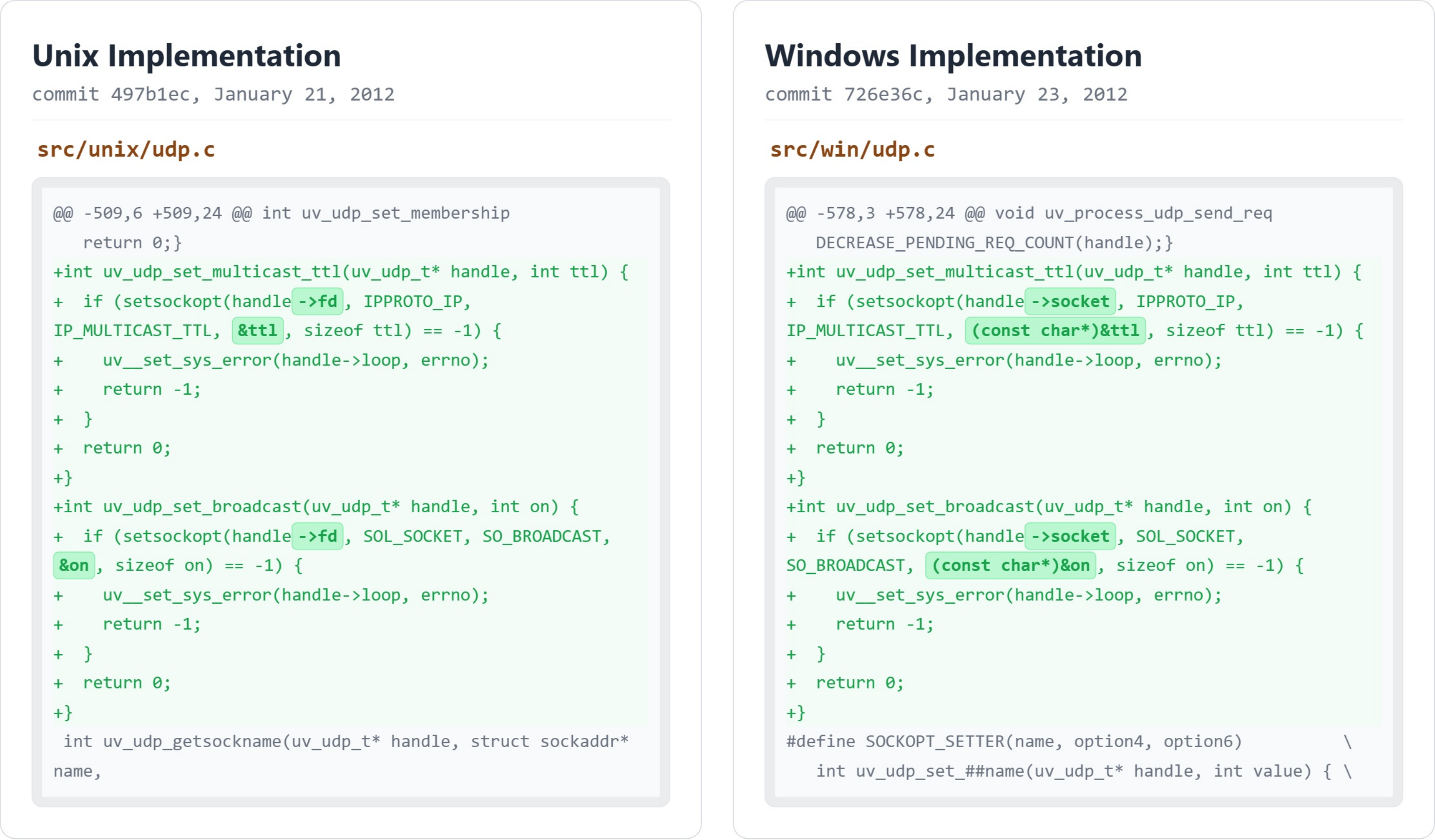}
\caption{\label{diff-example}A cross-OS code migration example from repository \textit{libuv/libuv}. The functionality was first implemented on Unix in an earlier commit and subsequently implemented on Windows in a later commit. During execution, each platform invokes only its corresponding platform-specific implementation.}
\end{figure*}

Our preliminary case studies illustrate how this challenge arises in real multi-platform repositories.
In these cases, functionality is implemented for one platform first, such as Linux, and later ported to another, such as Windows, using a separate platform-specific implementation.
Figure~\ref{diff-example} shows an example from the asynchronous I/O library \textit{libuv/libuv}.
Support for configuring UDP sockets for multicast and broadcast was first introduced in the Unix backend and was later added to the Windows backend in a separate commit.
The two implementations expose the same user-facing behavior, but they live in different code paths and rely on different operating-system interfaces.

Although such histories reveal migration work, migration tasks cannot be reliably extracted directly from individual commits.
A single migration may span multiple commits, and each commit may mix migration changes with unrelated bug fixes, refactoring, or feature development.
Using a commit as a task can therefore omit changes needed to complete the migration or include edits unrelated to it.
Constructing a coherent task from commit history therefore requires substantial manual effort to trace the migration across multiple commits and disentangle relevant edits from unrelated changes.

We address this challenge requiring heavy manual workload and automate the task-construction process by using differences in execution coverage to identify platform-specific implementations.
When a repository's tests are run on Linux and Windows, code associated with one backend is exercised on one OS but not the other.
We build \osbench{} around this observation using a differential-coverage pipeline.
We first build and test candidate repositories in native Linux and Windows environments, compare their coverage traces, and verify which differences reflect platform-specific behavior.
We then remove verified target-platform implementations to construct migration tasks, retaining tests that fail when the implementations are missing and regression tests that check unrelated behaviors are preserved.
This procedure creates controlled restoration tasks without requiring migration changes to align with historical commit boundaries.

\osbench{} contains 140 executable tasks from 111 open-source repositories spanning C, C++, C\#, Go, and Rust.
It includes both migration directions: 91 tasks migrate functionality supported on Windows to Linux, while 49 migrate Linux-supported functionality to Windows.
Each task provides a repository at a reproducible base commit, a natural-language migration objective, an executable environment, fail-to-pass tests for the target functionality, and pass-to-pass tests for regression checking.
We evaluate each patch on both the target and source platforms to check that it restores the missing functionality while preserving existing behavior under the test suites.

We evaluate six models using three Linux agent harnesses---SWE-agent, Claude Code, and OpenHands---as well as the Windows-native Win-agent harness.
The strongest configuration solves only 26.4\% of Linux-target tasks, while the best Windows configuration reaches 20.4\% on Windows-target tasks when success requires the patch to work correctly on both systems.
Performance also varies substantially across languages, migration directions, and agent harnesses.
In particular, changing the harness can noticeably change the performance of the same underlying model.
These results show that reliable migration remains difficult and that performance depends on both the model and the tools it uses to interact with the operating system.

Our main contributions are as follows:
\begin{itemize}
\item We formulate cross-OS code migration as an executable repository-level SWE task, where an agent must implement missing behavior on a target platform without breaking functionality on the source platform.
\item We introduce an LM-assisted, scalable task-construction pipeline that uses differential execution coverage between Linux and Windows to identify and verify platform-specific implementations without relying on historical commit boundaries.
\item We release \osbench{}, containing 140 validated tasks from 111 repositories across five programming languages and both Linux-to-Windows and Windows-to-Linux migration directions.
\item We evaluate six models with four agent harnesses and find low cross-platform success rates, with substantial variation across languages, migration directions, and agent designs.
\end{itemize}
\section{Related Work}

\subsection{SWE Benchmarks}

\textbf{General SWE benchmarks.}
Early code-generation benchmarks such as HumanEval~\citep{chen2021evaluating} and LiveCodeBench~\citep{jain2025livecodebench} largely focus on self-contained, function-level programming problems.
The SWE-bench family~\citep{jimenez2024swe,zhang2025swe,deng2025swe,jin2026chainswe} moves beyond this setting to repository-level issue resolution, where agents must identify relevant code, reason about dependencies, modify multiple files when necessary, and validate their changes in executable environments.
Agentic systems such as RepoLaunch~\citep{li2026repolaunch} further improve the scalability of constructing and reproducing these repository environments.
While SWE-bench primarily centers on resolving reported software issues, ProgramBench~\citep{yang2026programbench} evaluates whether language models can reconstruct complete software repositories from scratch.
These benchmarks typically execute and evaluate each task within a single OS environment, leaving the transfer of functionality across platform-specific implementations largely untested.

\textbf{Specialized SWE benchmarks.}
More recent benchmarks have started to focus on specific forms of software maintenance and evolution.
MigrationBench~\citep{liu2025migrationbench}, for example, evaluates repository-scale migrations between Java versions.
MigGPT~\citep{dang2026miggpt} studies compatibility changes required by third-party Linux libraries across Linux versions, while ChainSWE~\citep{jin2026chainswe} evaluates sequences of interdependent software engineering tasks that reflect how repositories evolve over time.
\osbench{} extends this direction to cross-OS migration, where an agent must translate functionality across different platform APIs and runtime behavior while preserving correctness on the original system.
To the best of our knowledge, \osbench{} is the first executable repository-level benchmark specifically designed for Linux--Windows code migration.

\subsection{SWE Agents}
SWE agents combine LMs with tools for navigating repositories, editing code, executing commands, and iterating on test results.
Open-source systems such as SWE-agent~\citep{yang2024swe, zhang2026sweedit} and OpenHands~\citep{wang2025openhands} provide agent--computer interfaces that allow models to inspect files, run shell commands, apply patches, and respond to execution feedback.
Commercial coding agents, including Codex~\citep{codex} and Claude Code~\citep{claudecode}, further support capabilities such as task planning, long-context repository analysis, and delegated subtasks/subagents.
More general computer-use agents, such as OpenClaw~\citep{openclaw}, extend this style of tool-mediated interaction beyond source-code repositories to broader OS workflows.

Despite recent progress, most SWE agent benchmarks and tool interfaces remain Linux-centric. 
Native Windows execution poses distinct challenges in shell semantics, path and escaping conventions, filesystem behavior, process management, compilation, and container support, affecting both code modification and agent–environment interaction. 
Recent incidents where coding agents unexpectedly deleted user data or exposed credentials in Windows environments further highlight the need for Windows-specific training, tooling, and evaluation. 
Existing benchmarks rarely assess cross-platform correctness: implementing platform-specific behavior on Windows while preserving Linux compatibility, or vice versa. 
\osbench{} addresses these gaps through native executable environments for both platforms, a Windows-oriented agent harness, and counter-platform regression testing.
\section{OS-Bench}
\label{sec:method}

\begin{figure*}[t]
    \centering
    \includegraphics[width=\textwidth]{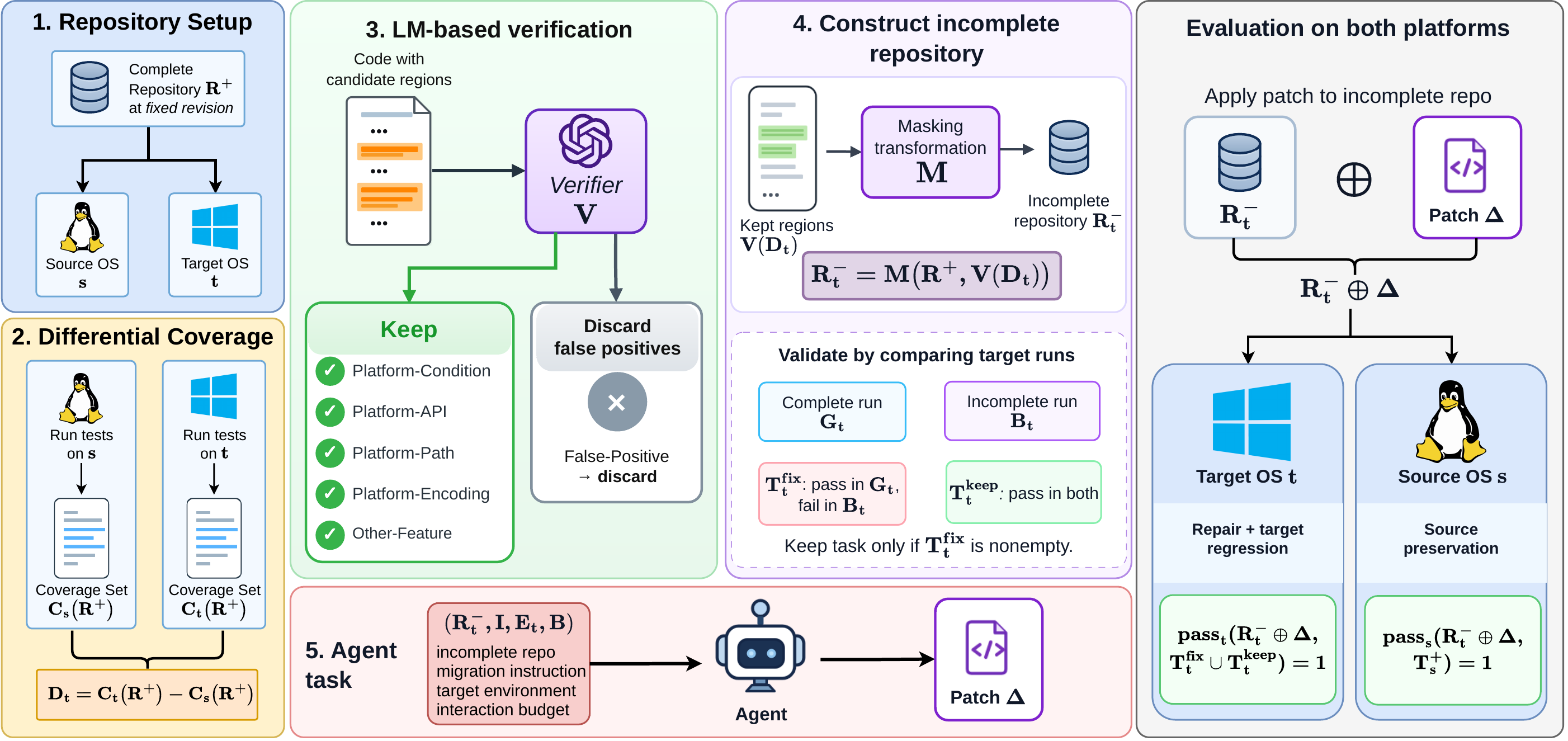}
    \caption{
    Overview of the \osbench{} construction and evaluation pipeline.
    }
    \label{fig:osbench-overview}
\end{figure*}

An \osbench{} task starts from a repository with part of its implementation for one operating system removed.
The agent restores support for that target system, and evaluation checks the submitted patch on both the target and source platforms.
Figure~\ref{fig:osbench-overview} summarizes the full \osbench{} construction and evaluation pipeline.

\subsection{Task Formulation}

Let $R^{+}$ denote the complete repository at a fixed revision, with executable test workloads on a source platform $s$ and a target platform $t$.
Let $C_s(R)$ and $C_t(R)$ be the sets of source lines executed by the repository's test workload on the two platforms.
We identify target-specific candidate code as
\begin{equation}
D_t = C_t(R^{+}) \setminus C_s(R^{+}).
\end{equation}
An LM-based verifier $V$ examines the source code and filters out coverage differences unrelated to platform-specific behavior.
A masking transformation $M$ then removes or replaces the verified target implementation:
\begin{equation}
R^{-}_t = M(R^{+}, V(D_t)).
\end{equation}

The agent receives $(R^{-}_t, I, E_t, B)$: the incomplete repository, a natural-language migration instruction, the target execution environment, and a fixed interaction budget.
It returns a patch $\Delta$.

Let $T_t^{\mathrm{fix}}$ denote target tests that do not pass in $R^{-}_t$ but pass in $R^{+}$, $T_t^{\mathrm{keep}}$ denote target tests that pass in both states, and $T_s^{+}$ denote tests that pass in the complete source environment.
A patch solves the task only if it satisfies both
\begin{align}
\operatorname{pass}_t(R^{-}_t \oplus \Delta,
T_t^{\mathrm{fix}} \cup T_t^{\mathrm{keep}}) &= 1, \\
\operatorname{pass}_s(R^{-}_t \oplus \Delta, T_s^{+}) &= 1.
\end{align}
The first condition checks target repair and regression tests; the second checks source-platform preservation.
We also require the source reference set $T_s^{+}$ to be nonempty.
The same patch is evaluated in both native environments.

\paragraph{Why test both platforms?}
A target test suite cannot directly exercise code compiled only on the source OS.
A patch can therefore pass the target suite while breaking a source-only branch or removing a guard needed by the source build.
Running the patch on both systems checks behavior that would otherwise go unobserved.
As with any test-based evaluation, passing these checks establishes correctness only for the behaviors they cover.

\subsection{Repository and Environment Selection}

We start from popular open-source repositories written in C, C++, C\#, Go, and Rust whose continuous-integration configurations indicate support for both Linux and Windows.
For each repository, RepoLaunch~\citep{li2026repolaunch} checks out a fixed revision, resolves dependencies, builds the project in a native environment, identifies a stable test command, and constructs a parser that converts raw test output into test names and statuses.
A repository is retained only when usable environments can be built on both operating systems.

The complete repository provides the reference behavior for task construction and evaluation.
The environments retain platform-native shells and toolchains, since path handling, command execution, and system APIs are part of the migration problem.

\subsection{Differential Coverage and Verification}
\label{sec:coverage}

We run each repository's tests with language-appropriate line-coverage instrumentation on both Linux and Windows.
Coverage paths are normalized to the repository root while accounting for differences such as platform-specific separators, generated build prefixes, Go module paths, and Rust registry paths.
For each file, we identify lines executed only on Linux or only on Windows and merge neighboring lines into candidate regions.

Coverage asymmetry alone does not establish that code is platform-specific.
A line may execute on only one system because of differences in test selection, nondeterminism, dependency versions, or unrelated feature flags.
We therefore use an LM-based verifier that receives the complete source file with Linux-only and Windows-only regions annotated and assigns each region one of six labels:
\begin{itemize}
\item \textsc{Platform-Conditional}: compile-time guards, build tags, platform-specific filenames, or runtime OS checks;
\item \textsc{Platform-API}: system calls, headers, commands, libraries, constants, or other facilities specific to one OS family;
\item \textsc{Platform-Path}: differences in separators, roots, drive prefixes, case sensitivity, or path-list semantics;
\item \textsc{Platform-Encoding}: line endings, locales, console code pages, Unicode conversion, or other platform-specific text representations;
\item \textsc{Other-Migration-Feature}: portability shims, dependency substitutions, capability probes, build adaptations, and related compatibility logic;
\item \textsc{False-Positive}: asymmetric execution with no inherent platform-specific semantics.
\end{itemize}
Regions assigned to the first five categories were retained while those labeled \textsc{False-Positive} were discarded.
The LM-retained lines were further filtered by human.
For each migration direction, the retained regions in a repository form one task.
A task may therefore involve several categories and require changes across multiple files.

\subsection{Constructing the Incomplete State}
\label{sec:construction}

For each retained repository, we first rebuild and test the complete version on the target platform and record its reference test status $G_t$.
An LM-based masker is prompted to substitute every verified target-only region with a placeholder which would be silently bypassed during execution, while avoiding inserting obvious exceptions or a replacement implementation to maintain a sufficiently high level of task difficulty.
Test files are excluded by default.
The LM-masked fields are manually verified to ensure that: (1) It masked all verified target-specific regions selected for the task with no exceptions or re-implementations inserted; and (2) It did not touch source-platform codes and source-platform test suite $T_s^{+}$ remains passed. 
The violations are manually fixed.

We then rebuild the incomplete repository and run the same target test suite to obtain status $B_t$.
The task labels are then derived directly from the difference between the complete and incomplete states:
\begin{align}
T_t^{\mathrm{fix}} &= \{x \mid G_t(x)=\mathrm{pass}
\land B_t(x)\neq\mathrm{pass}\}, \\
T_t^{\mathrm{keep}} &= \{x \mid G_t(x)=\mathrm{pass}
\land B_t(x)=\mathrm{pass}\}.
\end{align}
A candidate is kept only when $T_t^{\mathrm{fix}}$ is nonempty.
Thus every retained task has at least one previously passing test that fails, is skipped, or is absent after masking.
The reverse of the masking diff provides the reference (gold) repair patch.

To prevent recovery of the removed implementation from Git history, the original \emph{.git} data including the Git remote URL were removed and the repo is re-initialized with a root commit.

We assign each task a construction-time difficulty label based on the size of the verified migration region.
A task is \emph{hard} if the region exceeds 100 estimated lines or spans more than five files, \emph{medium} if it exceeds 30 lines or two files, and \emph{easy} otherwise.

\subsection{Benchmark Composition}

Table~\ref{tab:composition} summarizes the released benchmark.
\osbench{} contains 140 tasks drawn from 111 repositories, with 29 repositories contributing one task in both migration directions.
On average, each task spans seven files and 271 estimated platform-specific lines, while the largest task covers 152 files and 5,287 estimated lines.
Across all tasks, the gold patches add 56,908 lines and remove 14,459 lines.
The benchmark contains 44 easy, 38 medium, and 58 hard tasks.

\begin{table*}[t]
\centering
\small
\setlength{\tabcolsep}{5.9pt}
\caption{\label{tab:composition}
Composition of \osbench{}.
The first four columns report counts, while the last four report per-task
averages.
Fail-to-pass (F2P) checks pass in the complete target repository and stop
passing after masking; pass-to-pass (P2P) checks pass in both states.
Files and estimated lines of code (LoC) are computed over the verified
regions of each task; a file may therefore be counted in tasks from both
migration directions.}
\begin{tabular}{lrrrrrrrr}
\toprule
& \multicolumn{4}{c}{\textbf{Counts}}
& \multicolumn{4}{c}{\textbf{Per-task averages}} \\
\cmidrule(lr){2-5}
\cmidrule(lr){6-9}
Language
& Repos
& Tasks
& Win$\rightarrow$Linux
& Linux$\rightarrow$Win
& F2P checks
& P2P checks
& Files
& LoC \\
\midrule
C
& 2  & 2  & 1  & 1
& 21.5  & 12.5   & 2.0 & 73.0 \\

C++
& 6  & 7  & 6  & 1
& 48.9  & 57.0   & 9.7 & 669.3 \\

C\#
& 31 & 34 & 24 & 10
& 675.1 & 3101.4 & 6.2 & 99.0 \\

Go
& 28 & 38 & 20 & 18
& 509.8 & 624.8  & 9.0 & 324.3 \\

Rust
& 44 & 59 & 40 & 19
& 253.6 & 77.2   & 6.1 & 295.2 \\
\midrule
Overall
& 111 & 140 & 91 & 49
& 411.9 & 958.4 & 7.0 & 271.0 \\
\bottomrule
\end{tabular}
\end{table*}
\section{Experiments}
\label{sec:experiments}

We evaluate how reliably agents restore target-platform functionality while preserving source-platform behavior, and how success varies with migration direction, language, and agent harness.

\paragraph{Models and harnesses.}
We evaluate GPT-5.6-Sol, GPT-5.6-Terra, GPT-5.6-Luna, Claude-Opus-4.8, Claude-Sonnet-5, and DeepSeek-V4-Pro with medium thinking effort.
For the 91 Linux-target tasks, each model runs with Claude Code~\citep{claudecode}, OpenHands~\citep{wang2025openhands}, and SWE-agent~\citep{yang2024swe}.
As the three Linux-based agents currently do not support running in Windows docker containers, for the 49 Windows-target tasks, each model runs with Win-agent~\citep{li2026repolaunch}, a native Windows harness with a tool set aligned with that of Linux SWE-agent: file viewing, creation and editing, PowerShell command execution, and patch submission.
\Cref{appendix-agent-toolset} compares the tool sets of the four agent harnesses. 
Each run is limited to 100 agent steps and five hours.
The task instruction specifies the migration direction and asks the agent to fix compatibility issues and pass the target regression tests; each harness supplies its own workflow instructions.
\Cref{app:prompts} reproduces the prompt templates.

\paragraph{Evaluation.}
\label{sec:evaluation}
We apply each submitted patch to the incomplete repository and evaluate it in two stages.
Stage~I requires every target fail-to-pass and pass-to-pass check to pass.
Stage~II applies the same patch in the native source environment and requires every test that passed in the complete source repository to remain passing, with a nonempty reference set.
The primary metric is the percentage of tasks that pass both stages, using all 91 or 49 tasks in the corresponding split as the denominator.
Runs without a validated success contribute no successes.
\Cref{app:protocol} gives the full protocol.
\section{Results}
\label{sec:results}

\begin{table*}[t]
    \centering
    \small
    \caption{\label{cross-platform-success}
      Cross-platform success rates grouped by model. Linux evaluations use
      SWE-agent, Claude Code, and OpenHands, while Windows evaluations use
      Win-agent. Thinking effort is set to \textit{medium} for all models
      for a fair comparison.
    }
    \begin{tabular}{lllccccc}
      \hline
      \textbf{Model}
        & \textbf{Platform}
        & \textbf{Agent}
        & \textbf{C/C++}
        & \textbf{C\#}
        & \textbf{Go}
        & \textbf{Rust}
        & \textbf{Overall} \\
      \hline

      \multirow{4}{*}{\textbf{GPT-5.6-Sol}}
        & \multirow{3}{*}{Linux}
        & SWE-agent   & 0.0\% & 12.5\% & 5.0\%  & 32.5\% & 18.7\% \\
      & & Claude Code & 0.0\% & 12.5\% & 5.0\%  & 45.0\% & 24.2\% \\
      & & OpenHands  & 0.0\% & 12.5\% & 15.0\% & 32.5\% & 20.9\% \\
      \cline{2-8}
      & Windows & Win-agent
        & 0.0\% & 40.0\% & 5.6\% & 26.3\% & 20.4\% \\
      \hline

      \multirow{4}{*}{\textbf{GPT-5.6-Terra}}
        & \multirow{3}{*}{Linux}
        & SWE-agent   & 0.0\% & 12.5\% & 15.0\% & 32.5\% & 20.9\% \\
      & & Claude Code & 0.0\% & 12.5\% & 10.0\% & 35.0\% & 20.9\% \\
      & & OpenHands  & 0.0\% & 12.5\% & 15.0\% & 30.0\% & 19.8\% \\
      \cline{2-8}
      & Windows & Win-agent
        & 0.0\% & 30.0\% & 0.0\% & 10.5\% & 10.2\% \\
      \hline

      \multirow{4}{*}{\textbf{GPT-5.6-Luna}}
        & \multirow{3}{*}{Linux}
        & SWE-agent   & 0.0\% & 12.5\% & 0.0\%  & 17.5\% & 11.0\% \\
      & & Claude Code & 0.0\% & 8.3\%  & 15.0\% & 27.5\% & 17.6\% \\
      & & OpenHands  & 0.0\% & 8.3\%  & 15.0\% & 25.0\% & 16.5\% \\
      \cline{2-8}
      & Windows & Win-agent
        & 0.0\% & 20.0\% & 0.0\% & 0.0\% & 4.1\% \\
      \hline

      \multirow{4}{*}{\textbf{Claude-Opus-4.8}}
        & \multirow{3}{*}{Linux}
        & SWE-agent   & 0.0\% & 16.7\% & 15.0\% & 40.0\% & 25.3\% \\
      & & Claude Code & 0.0\% & 12.5\% & 10.0\% & 32.5\% & 19.8\% \\
      & & OpenHands  & 0.0\% & 12.5\% & 15.0\% & 45.0\% & 26.4\% \\
      \cline{2-8}
      & Windows & Win-agent
        & 50.0\% & 40.0\% & 5.6\% & 10.5\% & 16.3\% \\
      \hline

      \multirow{4}{*}{\textbf{Claude-Sonnet-5}}
        & \multirow{3}{*}{Linux}
        & SWE-agent   & 0.0\% & 0.0\%  & 0.0\%  & 25.0\% & 11.0\% \\
      & & Claude Code & 0.0\% & 4.2\%  & 5.0\%  & 32.5\% & 16.5\% \\
      & & OpenHands  & 0.0\% & 12.5\% & 10.0\% & 40.0\% & 23.1\% \\
      \cline{2-8}
      & Windows & Win-agent
        & 0.0\% & 0.0\% & 0.0\% & 0.0\% & 0.0\% \\
      \hline

      \multirow{4}{*}{\textbf{DeepSeek-V4-Pro}}
        & \multirow{3}{*}{Linux}
        & SWE-agent   & 0.0\% & 12.5\% & 15.0\% & 32.5\% & 20.9\% \\
      & & Claude Code & 0.0\% & 4.2\%  & 15.0\% & 32.5\% & 18.7\% \\
      & & OpenHands  & 0.0\% & 8.3\%  & 5.0\%  & 32.5\% & 17.6\% \\
      \cline{2-8}
      & Windows & Win-agent
        & 0.0\% & 20.0\% & 0.0\% & 0.0\% & 4.1\% \\
      \hline
    \end{tabular}
\end{table*}

\Cref{cross-platform-success} reports success when both target repair and source preservation are required, while \cref{app:single-platform} reports Stage~I success separately.

\subsection{Overall Performance and Source Preservation}

The strongest Linux configuration, Claude-Opus-4.8 with OpenHands, solves 24 of 91 tasks (26.4\%).
Claude-Opus-4.8 with SWE-agent follows at 25.3\%, while GPT-5.6-Sol with Claude Code reaches 24.2\%.
On Windows, GPT-5.6-Sol with Win-agent leads with 10 of 49 tasks solved (20.4\%), followed by Claude-Opus-4.8 at 16.3\%.
Thus, even the strongest configurations solve only about one quarter of Linux-target tasks and one fifth of Windows-target tasks.

Source-platform validation substantially changes the measured outcome.
For GPT-5.6-Sol with Claude Code, 50 Linux-target patches pass Stage~I, but only 22 pass both stages: success falls from 54.9\% to 24.2\%.
Of the 50 target-accepted patches, 28 (56.0\%) lack a validated source-platform pass.
For GPT-5.6-Sol with Win-agent, 13 Windows-target patches pass Stage~I and ten pass both stages (26.5\% versus 20.4\%).
Passing the target suite alone is therefore insufficient evidence of a successful migration.
\Cref{app:single-platform} gives the Stage-I-only results and their gaps from cross-platform success.

\subsection{Migration Direction and Programming Language}

Every model scores lower on Windows-target tasks than with its strongest Linux harness, and the best Windows score trails the best Linux score by 6.0 percentage points.
The model ranking also changes: GPT-5.6-Sol leads on Windows, whereas Claude-Opus-4.8 leads on Linux.
GPT-5.6-Terra reaches 19.8--20.9\% across Linux harnesses but 10.2\% on Windows.
Claude-Sonnet-5 reaches 23.1\% with OpenHands on Linux but solves no Windows tasks with Win-agent, while GPT-5.6-Luna and DeepSeek-V4-Pro each solve two Windows tasks (4.1\%).
These comparisons reflect the combined differences in task composition, model capability, and harness on the two different platforms.

Success is uneven across languages.
On Linux, Rust accounts for 18 of the 24 successes of Claude-Opus-4.8 with OpenHands and 18 of the 22 successes of GPT-5.6-Sol with Claude Code.
No Linux configuration solves a C/C++ task on both platforms, and Go success never exceeds 15.0\%.
On Windows, GPT-5.6-Sol and Claude-Opus-4.8 each solve four of ten C\# tasks (40.0\%), while their Go success is only one of 18 tasks (5.6\%).
GPT-5.6-Sol also leads on Windows Rust tasks, solving five of 19 (26.3\%).
The language profile therefore changes with the migration direction: Rust accounts for most Linux successes, whereas C\# contributes a larger share of Windows successes.
The small C/C++ groups---seven Linux tasks and two Windows tasks---limit broader conclusions about these languages.

\subsection{Sensitivity to the Agent Harness}

Harness choice also matters when the model and task split are held fixed.
Claude-Sonnet-5 ranges from ten Linux successes with SWE-agent (11.0\%) to 21 with OpenHands (23.1\%), an 11-task difference.
Claude-Opus-4.8 and GPT-5.6-Luna each show a six-task difference between their strongest and weakest Linux harnesses.
For Claude-Opus-4.8, OpenHands solves 24 tasks compared with 18 for Claude Code; for GPT-5.6-Luna, Claude Code solves 16 compared with ten for SWE-agent.
No harness is best for every model: OpenHands leads for Claude-Opus-4.8 and Claude-Sonnet-5, Claude Code leads for GPT-5.6-Sol and GPT-5.6-Luna, and SWE-agent leads for DeepSeek-V4-Pro.
GPT-5.6-Terra ties between SWE-agent and Claude Code.
The differences also depend on language: GPT-5.6-Sol with Claude Code solves more Rust tasks than with OpenHands (18 versus 13), whereas OpenHands solves more Go tasks (three versus one).
Model and harness should therefore be evaluated together.

Harness rankings can also change when source preservation is required.
For GPT-5.6-Luna, OpenHands has the higher target-only success rate (46.2\% versus 41.8\% with Claude Code), but Claude Code has the higher both-platform rate (17.6\% versus 16.5\%).
A harness that restores more target implementations can therefore produce fewer validated migrations.
This reversal reinforces the need to evaluate the same patches on both systems when comparing agent configurations.

\subsection{Failure Patterns}
\label{sec:failure-patterns}

\begin{table}[t]
\centering
\small
\caption{\label{tab:failure-categories}
Failure rates by migration category for GPT-5.6-Sol with Claude Code (Linux) and Win-agent (Windows).
Target is the percentage failing Stage~I; Both is the percentage failing the combined Stage~I--II criterion.
The denominator is the task count in each row and direction.
Labels overlap, so task counts should not be summed across rows.}
\begin{tabular}{lrrrcrrr}
\toprule
& \multicolumn{3}{c}{Windows$\rightarrow$Linux}
&& \multicolumn{3}{c}{Linux$\rightarrow$Windows} \\
\cmidrule(lr){2-4}\cmidrule(lr){6-8}
Category & Tasks & Target & Both && Tasks & Target & Both \\
\midrule
Platform conditionals & 71 & 43.7\% & 76.1\% && 38 & 68.4\% & 76.3\% \\
OS / library APIs     & 61 & 52.5\% & 78.7\% && 28 & 71.4\% & 75.0\% \\
Paths                 & 56 & 44.6\% & 78.6\% && 32 & 78.1\% & 81.2\% \\
Encoding              & 33 & 57.6\% & 81.8\% && 16 & 81.2\% & 93.8\% \\
Other compatibility   & 45 & 57.8\% & 82.2\% && 24 & 66.7\% & 70.8\% \\
\bottomrule
\end{tabular}
\end{table}

\begin{table}[t]
\centering
\small
\caption{\label{tab:failure-mechanisms}
Observed repair and verification patterns in the 35 recorded Stage~I failures for GPT-5.6-Sol with Claude Code on Linux.
Each share uses all 35 runs as its denominator.
Labels overlap: a run can exhibit multiple patterns, so the shares do not sum to 100\%.
Definitions are given in \cref{sec:failure-patterns}.}
\begin{tabular}{lrr}
\toprule
Failure pattern & Runs & Share \\
\midrule
Unrestored implementation & 18 & 51.4\% \\
Reduced verification scope & 19 & 54.3\% \\
Turn budget exhausted & 2 & 5.7\% \\
\bottomrule
\end{tabular}
\end{table}

\paragraph{Failures span platform behaviors.}
\Cref{tab:failure-categories} reports failure rates for GPT-5.6-Sol with Claude Code on Linux and Win-agent on Windows.
The categories overlap: a task involving both system APIs and path handling appears in both rows, so they describe task contents rather than independent causes of failure.
On Linux, 76.1--82.2\% of tasks in each category remain unsolved under the combined criterion.
For path-related tasks, the unsolved count increases from 25 of 56 under target-only evaluation to 44 of 56 when source preservation is required.
On Windows, 15 of 16 tasks involving encoding remain unsolved, as do 26 of 32 involving paths.
The difficulty therefore extends across conditional logic, APIs, paths, and encoding.

\paragraph{Partial repairs leave migration work incomplete.}
We audit the 35 Linux runs explicitly recorded as Stage~I failures for GPT-5.6-Sol with Claude Code.
This subset excludes incomplete or absent evaluations and patches that fail only at Stage~II.
\Cref{tab:failure-mechanisms} summarizes the audit; \emph{turn budget exhausted} denotes runs that reach their interaction limit.
In 18 of the 35 runs (51.4\%), at least one untouched production file retains missing behavior.
For example, in \texttt{yarp}, the agent restores certificate loading but leaves service registration incomplete: the reference repair also registers HTTP.sys delegation on Windows and a no-op fallback on Linux.
In \texttt{yek}, the agent adds Unix permission checks to handle tests run as a privileged user, but leaves the output-path logic incomplete.
When both an output directory and filename are supplied, the masked code still uses the filename alone rather than joining it with the directory.

\paragraph{Local verification can omit required checks.}
In 19 audited runs (54.3\%) in \Cref{tab:failure-mechanisms}, the agent's verification excludes tests, suites, or build modes required by the benchmark.
In \texttt{simdjson}, the agent disables single-header generation in its local build, although \texttt{simdjson-singleheader} is a required F2P check.
Its 108-test run therefore does not establish that the benchmark's required workload passes.
In \texttt{difftastic}, the final test output marks \texttt{test\_gzip\_is\_binary} as ignored, although this test belongs to the required F2P set and the benchmark explicitly runs ignored tests.
These omissions can coexist with incomplete repairs; the audit labels are not mutually exclusive.
The benchmark marks such runs unsuccessful when required checks do not pass, and the traces do not establish that the omissions were deliberate.
\FloatBarrier
\section{Conclusion}
We introduce \osbench{}, a benchmark of 140 code migration tasks between Linux and Windows.
Each task asks an agent to restore missing functionality on one platform while preserving behavior on the other.
The best evaluated configurations solve 26.4\% of Windows-to-Linux tasks and 20.4\% of Linux-to-Windows tasks under this requirement.
These results reveal that current agents still struggle with the demands of reliable cross-platform migration.
Closing this gap requires agents to track the full scope of a migration, restore platform-specific behavior across files, and validate changes on both operating systems.
\osbench{} provides a testbed for measuring progress toward these capabilities.

\section*{Limitations}
Our task-construction pipeline relies on cross-OS differences in execution coverage, so it cannot detect migration features implemented through cross-platform abstractions that follow the same code paths on both systems. 
Recovering such cases would require mining commit history to infer when and how cross-platform support was introduced. However, commits often mix unrelated changes and are rarely labeled as platform migrations, making similar large-scale automation like this work difficult and requiring substantial manual labeling effort. 
We therefore leave the use of historical signals to discover these additional migration patterns to future work.

\clearpage



\subsection*{Security Considerations}
\osbench{} evaluates target-platform repair and source-platform preservation only for behaviors covered by repository tests.
Prompt injection through poisoned repositories, including malicious comments or instructions, may cause unsafe code modifications or shell/PowerShell execution.
Migration patches may also introduce vulnerabilities despite passing tests on both platforms.
We recommend isolated, least-privilege execution with restricted network access and no sensitive data or production credentials.

\subsection*{Responsible AI}
Migration capabilities could be misused to adapt malicious software to additional platforms.
For production use, we recommend human review and approval of patches and high-risk actions, alongside security checks and regression testing on both platforms.
Developers and operators should retain responsibility for deployed changes, with execution logs and rollback procedures supporting oversight.

\bibliographystyle{unsrtnat}
\bibliography{references}

@inproceedings{jain2025livecodebench,
  title={Livecodebench: Holistic and contamination free evaluation of large language models for code},
  author={Jain, Naman and Gu, Alex and Li, Wen-Ding and Yan, Fanjia and Zhang, Tianjun and Wang, Sida and Solar-Lezama, Armando and Sen, Koushik and Stoica, Ion},
  booktitle={International Conference on Learning Representations},
  volume={2025},
  pages={58791--58831},
  year={2025}
}

@inproceedings{jimenez2024swe,
  title={Swe-bench: Can language models resolve real-world github issues?},
  author={Jimenez, Carlos E and Yang, John and Wettig, Alexander and Yao, Shunyu and Pei, Kexin and Press, Ofir and Narasimhan, Karthik},
  booktitle={International Conference on Learning Representations},
  volume={2024},
  pages={54107--54157},
  year={2024}
}

@misc{jin2026chainswe,
      title={ChainSWE: Benchmarking Coding Agents on Multi-Bug Software Maintenance}, 
      author={Qirui Jin and Lingching Tung and Kenan Li and Qiyang Shi and Yushi She and Huanzhong Jia and Harrison Zhao and Kejing Xia and Zhenbang Du and Jiaxin Pei and Zhenyu Zhang and Zhen Qi and Yuyan Duan and Wenke Lee and Zijian Jin},
      year={2026},
      eprint={2607.02606},
      archivePrefix={arXiv},
      primaryClass={cs.SE},
      url={https://arxiv.org/abs/2607.02606}, 
}

@misc{li2026repolaunch,
      title={RepoLaunch: Automating Build and Management of Code Repositories across Languages and Platforms}, 
      author={Kenan Li and Rongzhi Li and Linghao Zhang and Qirui Jin and Liao Zhu and Xiaosong Huang and Geng Zhang and Yikai Zhang and Shilin He and Chengxing Xie and Xin Zhang and Zijian Jin and Bowen Li and Chaoyun Zhang and Yu Kang and Yufan Huang and Elsie Nallipogu and Saravan Rajmohan and Qingwei Lin and Dongmei Zhang},
      year={2026},
      eprint={2603.05026},
      archivePrefix={arXiv},
      primaryClass={cs.SE},
      url={https://arxiv.org/abs/2603.05026}, 
}

@inproceedings{liu2025migrationbench,
author = {Liu, Linbo and Liu, Xinle and Zhou, Qiang and Chen, Lin and Liu, Yihan and Nguyen, Hoan and Tehrani, Behrooz Omidvar and Shen, Xi and Huan, Jun and Tripp, Omer and Deoras, Anoop},
title = {MigrationBench: Repository-Level Code Migration Benchmark from Java 8},
year = {2026},
isbn = {9798400722592},
publisher = {Association for Computing Machinery},
address = {New York, NY, USA},
url = {https://doi.org/10.1145/3770855.3817526},
doi = {10.1145/3770855.3817526},
booktitle = {Proceedings of the 32nd ACM SIGKDD Conference on Knowledge Discovery and Data Mining V.2},
pages = {9427–9438},
numpages = {12},
location = {Republic of Korea},
series = {KDD '26}
}

@article{dang2026miggpt,
  title={MigGPT: Harnessing large language models for automated migration of out-of-tree linux kernel patches across versions},
  author={Dang, Pucheng and Huang, Di and Li, Dong and Chen, Kang and Wen, Yuanbo and Guo, Qi and Hu, Xing},
  journal={Advances in Neural Information Processing Systems},
  volume={38},
  pages={38818--38851},
  year={2026}
}

@article{yang2024swe,
  title={Swe-agent: Agent-computer interfaces enable automated software engineering},
  author={Yang, John and Jimenez, Carlos E and Wettig, Alexander and Lieret, Kilian and Yao, Shunyu and Narasimhan, Karthik and Press, Ofir},
  journal={Advances in Neural Information Processing Systems},
  volume={37},
  pages={50528--50652},
  year={2024},
  url={https://openreview.net/forum?id=mXpq6ut8J3}
}

@inproceedings{
wang2025openhands,
title={OpenHands: An Open Platform for {AI} Software Developers as Generalist Agents},
author={Xingyao Wang and Boxuan Li and Yufan Song and Frank F. Xu and Xiangru Tang and Mingchen Zhuge and Jiayi Pan and Yueqi Song and Bowen Li and Jaskirat Singh and Hoang H. Tran and Fuqiang Li and Ren Ma and Mingzhang Zheng and Bill Qian and Daniel Shao and Niklas Muennighoff and Yizhe Zhang and Binyuan Hui and Junyang Lin and Robert Brennan and Hao Peng and Heng Ji and Graham Neubig},
booktitle={The Thirteenth International Conference on Learning Representations},
volume={2025},
pages={65882--65919},
year={2025},
url={https://openreview.net/forum?id=OJd3ayDDoF}
}

@article{zhang2025swe,
  title={SWE-bench Goes Live!},
  author={Linghao Zhang and Shilin He and Chaoyun Zhang and Yu Kang and Bowen Li and Chengxing Xie and Junhao Wang and Maoquan Wang and Yufan Huang and Shengyu Fu and Elsie Nallipogu and Qingwei Lin and Yingnong Dang and Saravan Rajmohan and Dongmei Zhang},
  journal={Advances in Neural Information Processing Systems},
  volume={38},
  year={2025}
}

@misc{zhang2026sweedit,
      title={SWE-Edit: Rethinking Code Editing for Efficient SWE-Agent}, 
      author={Yikai Zhang and Jiaxin Pei and Kenan Li and Qirui Jin and Maoquan Wang and Jin Pan and Yu Kang and Shengyu Fu and Elsie Nallipogu and Junjie Hu and Yufan Huang and Zijian Jin},
      year={2026},
      eprint={2604.26102},
      archivePrefix={arXiv},
      primaryClass={cs.SE},
      url={https://arxiv.org/abs/2604.26102}, 
}

@misc{li2026oracle,
      title={ORACLE-SWE: Quantifying the Contribution of Oracle Information Signals on SWE Agents}, 
      author={Kenan Li and Qirui Jin and Liao Zhu and Xiaosong Huang and Yijia Wu and Yikai Zhang and Xin Zhang and Zijian Jin and Yufan Huang and Elsie Nallipogu and Chaoyun Zhang and Yu Kang and Saravan Rajmohan and Qingwei Lin and Wenke Lee and Dongmei Zhang},
      year={2026},
      eprint={2604.07789},
      archivePrefix={arXiv},
      primaryClass={cs.MA},
      url={https://arxiv.org/abs/2604.07789}, 
}

@misc{yang2026programbench,
    title={ProgramBench: Can Language Models Rebuild Programs From Scratch?},
    author={John Yang and Kilian Lieret and Jeffrey Ma and Parth Thakkar and Dmitrii Pedchenko and Sten Sootla and Emily McMilin and Pengcheng Yin and Rui Hou and Gabriel Synnaeve and Diyi Yang and Ofir Press},
    year={2026},
    eprint={2605.03546},
    archivePrefix={arXiv},
    primaryClass={cs.SE},
    url={https://arxiv.org/abs/2605.03546},
}

@misc{deng2025swe,
      title={SWE-Bench Pro: Can AI Agents Solve Long-Horizon Software Engineering Tasks?}, 
      author={Xiang Deng and Jeff Da and Edwin Pan and Yannis Yiming He and Charles Ide and Kanak Garg and Niklas Lauffer and Andrew Park and Nitin Pasari and Chetan Rane and Karmini Sampath and Maya Krishnan and Srivatsa Kundurthy and Sean Hendryx and Zifan Wang and Vijay Bharadwaj and Jeff Holm and Raja Aluri and Chen Bo Calvin Zhang and Noah Jacobson and Bing Liu and Brad Kenstler},
      year={2025},
      eprint={2509.16941},
      archivePrefix={arXiv},
      primaryClass={cs.SE},
      url={https://arxiv.org/abs/2509.16941}, 
}

@misc{chen2021evaluating,
      title={Evaluating Large Language Models Trained on Code}, 
      author={Mark Chen and Jerry Tworek and Heewoo Jun and Qiming Yuan and Henrique Ponde de Oliveira Pinto and Jared Kaplan and Harri Edwards and Yuri Burda and Nicholas Joseph and Greg Brockman and Alex Ray and Raul Puri and Gretchen Krueger and Michael Petrov and Heidy Khlaaf and Girish Sastry and Pamela Mishkin and Brooke Chan and Scott Gray and Nick Ryder and Mikhail Pavlov and Alethea Power and Lukasz Kaiser and Mohammad Bavarian and Clemens Winter and Philippe Tillet and Felipe Petroski Such and Dave Cummings and Matthias Plappert and Fotios Chantzis and Elizabeth Barnes and Ariel Herbert-Voss and William Hebgen Guss and Alex Nichol and Alex Paino and Nikolas Tezak and Jie Tang and Igor Babuschkin and Suchir Balaji and Shantanu Jain and William Saunders and Christopher Hesse and Andrew N. Carr and Jan Leike and Josh Achiam and Vedant Misra and Evan Morikawa and Alec Radford and Matthew Knight and Miles Brundage and Mira Murati and Katie Mayer and Peter Welinder and Bob McGrew and Dario Amodei and Sam McCandlish and Ilya Sutskever and Wojciech Zaremba},
      year={2021},
      eprint={2107.03374},
      archivePrefix={arXiv},
      primaryClass={cs.LG},
      url={https://arxiv.org/abs/2107.03374}, 
}

@misc{claudecode,
  title        = {Claude Code},
  author       = {Anthropic},
  howpublished = {\url{https://code.claude.com/}},
  year         = {2025}
}

@misc{codex,
  title        = {Codex},
  author       = {OpenAI},
  howpublished = {\url{https://openai.com/codex/}},
  year         = {2025}
}

@misc{openclaw,
  title        = {OpenClaw : The AI that really does things},
  author       = {Peter Steinberger},
  howpublished = {\url{https://github.com/openclaw/openclaw}},
  year         = {2026}
}

\newpage
\appendix
\section{Detailed Experiments}
\label{app:protocol}

\subsection{Budgets and Execution Environments}

Each run receives a limit of 100 agent steps or model calls, according to the harness, with a five-hour wall-clock limit; SWE-agent stops at its 100-call limit.
Linux agents work in native Linux environments, and Win-agent uses a native Windows environment with PowerShell.
The prompts specify the migration direction and target-platform test objective, while the evaluator independently checks source-platform preservation.
The wrappers in \cref{app:prompts} also constrain test-file edits and Git use.

\subsection{Evaluation Stages}
\label{app:evaluation-stages}

\paragraph{Stage I: target-platform repair.}
We apply the submitted patch to the incomplete repository, run the native build and test commands, and parse the output into individual test results.
Every required fail-to-pass (F2P) and pass-to-pass (P2P) check must be recorded as passing.
Let $\operatorname{status}_p(x;\Delta)$ denote the recorded status of test $x$ on platform $p$ after applying patch $\Delta$.
The target check is
\begin{equation}
S_t(\Delta) =
\mathbb{1}\!\left[
\bigwedge_{x \in T_t^{\mathrm{fix}} \cup T_t^{\mathrm{keep}}}
\operatorname{status}_t(x;\Delta)=\mathrm{pass}\right].
\end{equation}
A compilation failure, timeout, missing required test result, or failed required test makes the run unsuccessful.

\paragraph{Stage II: source-platform preservation.}
Patches that pass Stage~I are tested on the source platform.
We first run the complete repository in the native source environment and collect its passing tests in the reference set $T_s^{+}$.
We then reverse the gold repair patch to recreate the incomplete repository, apply the agent's patch unchanged, and rerun the same tests.
Stage~II passes only if the reference set is nonempty and every test in it still passes:
\begin{equation}
S_s(\Delta) =
\mathbb{1}\!\left[
|T_s^{+}|>0 \land
\bigwedge_{x \in T_s^{+}} \operatorname{status}_s(x;\Delta)=\mathrm{pass}\right].
\end{equation}
Tests that already failed in the complete source repository are excluded.
A migration is solved only when both stages pass:
\begin{equation}
S(\Delta)=S_t(\Delta)\cdot S_s(\Delta).
\end{equation}

\section{Agent Harness Tool Set Comparison}
\label{appendix-agent-toolset}

\Cref{tab:toolset-comparison} compares the tool sets of the four agent harnesses under evaluation.

\begin{table}[htbp]
\centering
\small
\caption{\label{tab:toolset-comparison}
Comparison of the tool sets of the four agent harnesses under evaluation.}
\begin{tabular}{lcccc}
\toprule
Agent Harness & Win-agent & SWE-agent & OpenHands & ClaudeCode \\
\midrule
Supported docker container type & Linux\&Windows & Linux & Linux & Linux  \\
File view/create/edit & \checkmark & \checkmark & \checkmark & \checkmark \\
Shell command execution & \checkmark (Powershell) & \checkmark (Unix Bash) & \checkmark (Unix Bash) & \checkmark (Unix Bash) \\
Task planning & × & × & \checkmark & \checkmark \\
Sub-agent delegation & × & × & × & \checkmark \\
\bottomrule
\end{tabular}
\end{table}

\section{Single-Platform Success Rates}
\label{app:single-platform}
\label{app:aggregation}

Single-platform success measures target repair alone: every required target F2P and P2P check must pass, without imposing the source-platform check.
It uses the same patches, budgets, task splits, and denominators as the main evaluation.
\Cref{tab:single-platform} reports the number and percentage of accepted tasks under this criterion alongside success on both platforms.
The gap is computed from the difference in accepted task counts before rounding, in percentage points.
Claude-Sonnet-5 uses the union of two runs under each criterion, counting each task at most once.
A both-platform success requires the same patch to pass both stages.

\begin{table}[htbp]
\centering
\small
\setlength{\tabcolsep}{4pt}
\caption{\label{tab:single-platform}
Single-platform (Stage~I) and cross-platform (Stages~I--II) success.
Cells give accepted tasks and percentages, with denominators of 91
for Linux and 49 for Windows. Gap is the difference in percentage
points computed before rounding. Claude-Sonnet-5 pools two runs;
see \cref{app:aggregation}.}
\begin{tabular}{lllrrr}
\toprule
Model & Target & Agent & Target only & Both platforms & Gap \\
\midrule
GPT-5.6-Sol & Linux & SWE-agent & 43 (47.3\%) & 17 (18.7\%) & 28.6 \\
 & Linux & Claude Code & 50 (54.9\%) & 22 (24.2\%) & 30.8 \\
 & Linux & OpenHands & 43 (47.3\%) & 19 (20.9\%) & 26.4 \\
 & Windows & Win-agent & 13 (26.5\%) & 10 (20.4\%) & 6.1 \\
\midrule
GPT-5.6-Terra & Linux & SWE-agent & 40 (44.0\%) & 19 (20.9\%) & 23.1 \\
 & Linux & Claude Code & 42 (46.2\%) & 19 (20.9\%) & 25.3 \\
 & Linux & OpenHands & 45 (49.5\%) & 18 (19.8\%) & 29.7 \\
 & Windows & Win-agent & 11 (22.4\%) & 5 (10.2\%) & 12.2 \\
\midrule
GPT-5.6-Luna & Linux & SWE-agent & 20 (22.0\%) & 10 (11.0\%) & 11.0 \\
 & Linux & Claude Code & 38 (41.8\%) & 16 (17.6\%) & 24.2 \\
 & Linux & OpenHands & 42 (46.2\%) & 15 (16.5\%) & 29.7 \\
 & Windows & Win-agent & 6 (12.2\%) & 2 (4.1\%) & 8.2 \\
\midrule
Claude-Opus-4.8 & Linux & SWE-agent & 49 (53.8\%) & 23 (25.3\%) & 28.6 \\
 & Linux & Claude Code & 39 (42.9\%) & 18 (19.8\%) & 23.1 \\
 & Linux & OpenHands & 50 (54.9\%) & 24 (26.4\%) & 28.6 \\
 & Windows & Win-agent & 11 (22.4\%) & 8 (16.3\%) & 6.1 \\
\midrule
Claude-Sonnet-5 & Linux & SWE-agent & 22 (24.2\%) & 10 (11.0\%) & 13.2 \\
 & Linux & Claude Code & 41 (45.1\%) & 15 (16.5\%) & 28.6 \\
 & Linux & OpenHands & 44 (48.4\%) & 21 (23.1\%) & 25.3 \\
 & Windows & Win-agent & 4 (8.2\%) & 0 (0.0\%) & 8.2 \\
\midrule
DeepSeek-V4-Pro & Linux & SWE-agent & 36 (39.6\%) & 19 (20.9\%) & 18.7 \\
 & Linux & Claude Code & 32 (35.2\%) & 17 (18.7\%) & 16.5 \\
 & Linux & OpenHands & 35 (38.5\%) & 16 (17.6\%) & 20.9 \\
 & Windows & Win-agent & 4 (8.2\%) & 2 (4.1\%) & 4.1 \\
\bottomrule
\end{tabular}
\end{table}

\paragraph{Target success substantially exceeds migration success.}
For GPT-5.6-Sol with Claude Code, target-only evaluation accepts 50 of 91 Linux tasks (54.9\%), while both-platform evaluation accepts 22 (24.2\%).
The 28-task difference is 30.8 percentage points, and 56.0\% of target-accepted tasks lack a validated source-platform pass.
On Windows, the same model with Win-agent drops from 13 of 49 tasks (26.5\%) to ten (20.4\%), a 6.1-point gap.
These differences measure how much target-only evaluation overstates the number of validated migrations; they do not by themselves identify the cause of each unsuccessful source evaluation.

\paragraph{The gap depends on the harness.}
For GPT-5.6-Luna on Linux, OpenHands has the higher target-only rate (46.2\% versus 41.8\% with Claude Code), but Claude Code has the higher both-platform rate (17.6\% versus 16.5\%).
Thus, a harness ranking based on target repair can change when source preservation is required.
Across the reported configurations, Linux target-only rates range from 22.0\% to 54.9\%, compared with 11.0--26.4\% on both platforms.
On Windows, the corresponding ranges are 8.2--26.5\% and 0.0--20.4\%.

\section{Prompts}
\label{app:prompts}

This section reproduces the benchmark-specific prompt templates in the construction pipeline and evaluation harnesses.

\subsection{Platform-Specific Region Verification}

\begin{tcolorbox}[colback=blue!3, colframe=blue!50, title=Verification Prompt, breakable, fontupper=\scriptsize]
\begin{lstlisting}[style=patchbox]
You are a senior cross-platform migration specialist with deep expertise in {language} platform-specific code. Your task is to classify code regions that execute on only one platform (Linux or Windows) and determine whether they represent genuine platform dependencies or false positives.

## Source File: {filepath}

The complete source file is shown below with annotations:
- [L] = Line executes ONLY on Linux (not on Windows)
- [W] = Line executes ONLY on Windows (not on Linux)
- (no marker) = Line executes on both platforms or is not in coverage

```{language_ext}
{annotated_source}
```

## Coverage Analysis

Linux-only line ranges (execute on Linux, NOT Windows):
{linux_only_summary}

Windows-only line ranges (execute on Windows, NOT Linux):
{windows_only_summary}

## Classification Taxonomy

### Category Definitions ###

=== GENERAL CLASS DEFINITIONS ===

>>> PLATFORM_CONDITIONAL
  Code whose execution or compilation is selected by an explicit platform
  condition. This includes compile-time guards, build tags, platform-specific
  filenames, preprocessor branches, attributes, and runtime OS checks. Classify
  guarded code here even when the guarded statements are otherwise generic.

>>> PLATFORM_API
  Code that calls, imports, defines, wraps, or configures APIs, constants,
  commands, headers, crates, packages, namespaces, or system facilities that are
  specific to one OS family or behave incompatibly across OS families.

>>> PLATFORM_PATH
  Code that handles platform-dependent filesystem path syntax or semantics,
  including separators, drive prefixes, device names, root conventions, home or
  temp locations, path-list separators, and case-sensitivity behavior.

>>> PLATFORM_ENCODING
  Code that handles platform-dependent text, byte, locale, terminal, argument,
  environment, or line-ending representations, including Unicode conversion and
  console/codepage behavior.

>>> OTHER_MIGRATION_FEATURE
  Other platform-only code related to cross-platform migration or compatibility
  that does not fit the conditional/API/path/encoding classes. Use this for
  compatibility shims, fallback behavior, capability probing, dependency
  substitution, build/test adaptation, permission/process/filesystem behavior
  normalization, or other portability work needed for equivalent behavior across
  Linux and Windows.

>>> FALSE_POSITIVE
  Coverage differs between platforms, but the code is not inherently
  platform-specific. Use this for auto-generated files, test selection, timing, 
  dependency/version skew, unrelated feature flags, incidental coverage gaps, 
  or generic logic with no platform guard/API/path/encoding semantics.


### Language-specific Examples ###
{language_taxonomy}

## Task

For EACH contiguous block of platform-only lines listed above, classify it.

## IMPORTANT INSTRUCTIONS:
1. Analyze EVERY platform-only line block. Do not skip any.
2. Use the FULL file context to understand each region's purpose.
3. Code in directories like Interop/Linux/, Interop/Windows/, platform/, unix/, win32/ is platform-specific by definition.
4. Consider the SURROUNDING code (guards, conditionals, includes) not just the marked lines themselves.
5. When a marked line is inside a platform-guarded block (e.g., inside #ifdef _WIN32 or #[cfg(unix)]), the entire block is platform-specific even if only some lines are marked.
6. Do NOT classify based on regex pattern matching alone. Use your understanding of what the code actually does.
7. For each region, also identify:
  - functional_area: the precise functionality this code provides (e.g., "POSIX symlink resolution", "sigaction handler setup", "terminal raw mode control", "Windows path separator normalization")
  - feature_group: a short snake_case key for grouping regions that require the same exact platform knowledge across files. Do NOT use broad same-type labels such as "posix_filesystem" when regions involve different knowledge like symlink resolution, directory iteration, chmod permissions, path separators, or timestamp semantics. Prefer precise keys such as "posix_symlink_resolution", "posix_directory_iteration", "windows_registry_lookup", "sigaction_handler_setup".
8. Return a JSON array with one entry per contiguous platform-only line block.
  - If Linux-only and Windows-only blocks are adjacent, still return them as
separate entries with the correct `platform` value.
9. If several adjacent discrete line numbers (such as {33, 35, 37, 40, 42}) are all
windows_only or linux_only, with no counterpart mixed in, and they either all belong to platform-specific or
all belong to non-platform-specific, then consider merging them into a single entry: [start_line=33, end_line=42]
in your answer list to compact your answer.
10. If there are too many diff lines so that your output is too long,
only output entries that are real platform-specific lines and ignore FALSE_POSITIVE lines in your answer.
The only requirement is that all the real platform-specific lines must be recorded in your answer.


Now, output a list[dict] in JSON format inside a ```json``` fence:
```json
[
  {
    "start_line": <int, boundary included>,
    "end_line": <int, boundary included>,
    "platform": "linux_only|windows_only",
    "is_platform_specific": <bool>,
    "classification": "PLATFORM_CONDITIONAL|PLATFORM_API|PLATFORM_PATH|PLATFORM_ENCODING|OTHER_MIGRATION_FEATURE|FALSE_POSITIVE",
    "functional_area": "<what this code does>",
    "feature_group": "<snake_case grouping key>",
    "explanation": "<cite specific APIs, guards, or patterns that led to this classification>"
  },
  {
    "start_line": <int>,
    "end_line": <int>,
    "platform": "linux_only|windows_only",
    "is_platform_specific": <bool>,
    "classification": "PLATFORM_CONDITIONAL|PLATFORM_API|PLATFORM_PATH|PLATFORM_ENCODING|OTHER_MIGRATION_FEATURE|FALSE_POSITIVE",
    "functional_area": "<what this code does>",
    "feature_group": "<snake_case grouping key>",
    "explanation": "<cite specific APIs, guards, or patterns that led to this classification>"
  },
  ......
]
```

For example:
```json
[
  {
    "start_line": 131,
    "end_line": 138,
    "platform": "windows_only",
    "is_platform_specific": false,
    "classification": "FALSE_POSITIVE",
    "functional_area": "terminal color reset dispatch",
    "feature_group": "termcolor_stream_wrapper",
    "explanation": "This block implements `WriteColor::reset` via straightforward delegation to the wrapped writer. There are no Windows-specific imports, cfg attributes, path semantics, or encoding handling in this code, so the Windows-only execution is an incidental coverage artifact."
  },
  {
    "start_line": 165,
    "end_line": 169,
    "platform": "windows_only",
    "is_platform_specific": false,
    "classification": "FALSE_POSITIVE",
    "functional_area": "unit testing of invalid utf8 byte patterns",
    "feature_group": "pattern_utf8_validation_tests",
    "explanation": "This test validates pattern_from_bytes using an invalid UTF-8 byte sequence. It is not platform-specific; it just happened to run only on Windows in the coverage data. The actually platform-conditional test in this file is the separate #[cfg(unix)] os() test, not this one."
  }
]
```

Return ONLY the JSON array, no commentary before or after.
\end{lstlisting}
\end{tcolorbox}

\paragraph{Language-specific taxonomy blocks.}
Only the block matching the source file's language is inserted in each verifier call; C and C++ share a block.

\begin{tcolorbox}[colback=blue!3, colframe=blue!50, title=Language Taxonomy Prompt, breakable, fontupper=\scriptsize]
\begin{lstlisting}[style=patchbox]
=== C/C++ PLATFORM-SPECIFIC TAXONOMY ===

>>> PLATFORM_CONDITIONAL -- Compile-time platform checks:
  #ifdef _WIN32, #ifdef __linux__, #ifdef __unix__, #ifdef __APPLE__
  #ifdef _MSC_VER (MSVC compiler), #ifdef __GNUC__ (GCC)
  #if defined(_WIN32), #if defined(__linux__)
  #ifndef _WIN32, #if !defined(_WIN32)
  #ifdef WINAPI, #ifdef _WINDOWS

>>> PLATFORM_API -- OS-specific API calls:
  Windows: CreateFile, ReadFile, WriteFile, CloseHandle
  Windows: GetLastError, FormatMessage, SetLastError
  Windows: WSAStartup, WSACleanup, SOCKET, WSADATA
  Windows: CreateProcess, WaitForSingleObject, CreateThread
  Windows: RegOpenKeyEx, RegQueryValueEx (Registry)
  Windows: LoadLibrary, GetProcAddress
  Linux/POSIX: fork(), exec*(), waitpid(), kill()
  Linux/POSIX: open(), read(), write(), close(), ioctl()
  Linux/POSIX: pthread_*, sigaction, signal()
  Linux/POSIX: mmap, mprotect, sysconf
  Headers: <windows.h>, <winsock2.h>, <unistd.h>, <sys/socket.h>

>>> PLATFORM_PATH -- File path and filesystem differences:
  Path separator \ vs /
  Drive letters (C:\, D:\)
  MAX_PATH vs PATH_MAX
  Case sensitivity in filenames
  Symlink handling differences

>>> PLATFORM_ENCODING -- Encoding/locale differences:
  Wide char (wchar_t) on Windows vs UTF-8 on Linux
  _wfopen vs fopen, wmain vs main
  Line ending \r\n vs \n
  Codepage handling

>>> OTHER_MIGRATION_FEATURE -- Other migration/compatibility features:
  Cross-platform compatibility wrappers or fallback implementations
  CMake/autotools/MSBuild source selection or configuration for portability
  Normalizing permissions, process, thread, terminal, or filesystem behavior
  Test helpers or adapters that validate equivalent behavior across platforms

>>> FALSE_POSITIVE -- Not platform-specific:
  Test infrastructure differences
  Compiler optimization differences affecting coverage
  Header inclusion order differences
  Googletest/CTest framework noise
  Standard library implementation differences (libc vs MSVCRT)
  that don't affect API behavior

=== C# PLATFORM-SPECIFIC TAXONOMY ===

>>> PLATFORM_CONDITIONAL -- Runtime platform checks:
  RuntimeInformation.IsOSPlatform(OSPlatform.Windows/Linux/OSX)
  OperatingSystem.IsWindows(), OperatingSystem.IsLinux(), OperatingSystem.IsMacOS()
  [SupportedOSPlatform("windows")], [UnsupportedOSPlatform("windows")]
  #if WINDOWS, #if LINUX
  PlatformID enum checks

>>> PLATFORM_API -- Windows-specific APIs requiring migration:
  Registry: Microsoft.Win32.Registry, RegistryKey, HKEY_*
  WMI: System.Management.*, ManagementObject, ManagementObjectSearcher
  COM: [ComImport], Type.GetTypeFromProgID, Marshal.ReleaseComObject
  GDI+: System.Drawing.* (Bitmap, Graphics, Image, Font) on .NET 6+
  ETW: TraceEventSession, ETWTraceEventSource
  Events: EventLog, PerformanceCounter, ServiceController, ServiceBase
  P/Invoke: [DllImport], kernel32, user32, advapi32, ntdll
  Identity: WindowsIdentity, WindowsPrincipal, LogonUser
  AD: System.DirectoryServices (non-LDAP)
  Audio: System.Media.SoundPlayer, NAudio
  Certificates: X509Store+StoreLocation.LocalMachine (Windows cert store)

>>> PLATFORM_PATH -- File path differences:
  Hardcoded C:\, D:\ paths, backslash-only path literals
  Path.DirectorySeparatorChar when used in platform-specific logic
  Environment.GetFolderPath with Windows-only SpecialFolder values

>>> PLATFORM_ENCODING -- Encoding differences:
  Console.OutputEncoding platform differences
  Line ending handling

>>> OTHER_MIGRATION_FEATURE -- Other migration/compatibility features:
  Cross-platform compatibility adapters or fallback implementations
  Target framework/runtime compatibility code for OS behavior parity
  Normalizing services, permissions, process, terminal, or environment behavior
  Test helpers or adapters that validate equivalent behavior across platforms

>>> FALSE_POSITIVE -- Not platform-specific:
  #if NETFRAMEWORK / #if NET6_0 (framework version, NOT platform)
  Standard BCL (System.IO.Path, System.Net.Http, System.Text.Json, LINQ)
  Async patterns (Task, async/await, CancellationToken)
  Modern frameworks (ASP.NET Core, EF Core, gRPC, HttpClient)
  Test infrastructure differences
  Auto generated files that are not actually source code written by human

=== GO PLATFORM-SPECIFIC TAXONOMY ===

>>> PLATFORM_CONDITIONAL -- Compile-time or runtime platform checks:
  runtime.GOOS == "linux" / "windows" / "darwin"
  //go:build linux, //go:build windows, //go:build unix
  // +build linux, // +build windows (legacy build tags)
  Files named *_linux.go, *_windows.go, *_darwin.go, *_unix.go
  Build constraints in file names (platform-specific file selection)

>>> PLATFORM_API -- OS-specific API calls:
  syscall.* (Syscall, RawSyscall, platform-specific constants)
  golang.org/x/sys/unix.*, golang.org/x/sys/windows.*
  os/signal handling differences
  os/exec with platform-specific commands
  os.Getenv("HOME") vs os.Getenv("USERPROFILE")
  net.Listen with platform-specific socket options

>>> PLATFORM_PATH -- File path and filesystem differences:
  filepath.Separator, os.PathSeparator, os.PathListSeparator
  filepath.ToSlash, filepath.FromSlash
  filepath.VolumeName (Windows drive letters)
  Hardcoded / vs \ in path construction
  Case sensitivity in file lookups

>>> PLATFORM_ENCODING -- Encoding/locale differences:
  Line ending normalization
  Console encoding/codepage differences
  Locale-dependent string operations

>>> OTHER_MIGRATION_FEATURE -- Other migration/compatibility features:
  Cross-platform compatibility wrappers or fallback implementations
  Build/module integration for target-specific dependencies
  Normalizing environment, permissions, process, terminal, or filesystem behavior
  Test helpers or adapters that validate equivalent behavior across platforms

>>> FALSE_POSITIVE -- Not platform-specific:
  Test infrastructure differences (test runner behavior)
  Environment variable differences unrelated to platform
  Timing-dependent coverage (goroutine scheduling)
  Template library functions that work identically on all platforms
  Pure computation / string manipulation that happens to have different coverage

=== RUST PLATFORM-SPECIFIC TAXONOMY ===

>>> PLATFORM_CONDITIONAL -- Compile-time or runtime platform checks:
  #[cfg(unix)], #[cfg(windows)], #[cfg(target_os = "linux")]
  #[cfg(target_family = "unix")], #[cfg(target_family = "windows")]
  cfg!(unix), cfg!(windows), cfg!(target_os = ...)
  #[cfg(not(windows))], #[cfg(any(unix, target_os = ...))]

>>> PLATFORM_API -- OS-specific API calls:
  std::os::unix::*, std::os::windows::*
  libc::* (fork, exec*, signal, pipe, ioctl, mmap, mprotect)
  winapi::*, windows_sys::*, windows::Win32::*
  nix::* (nix crate for Unix APIs)
  std::os::unix::fs::symlink, std::os::unix::fs::PermissionsExt
  std::os::windows::ffi::OsStrExt, OsStringExt
  std::process::Command with platform-specific args
  Signal handling (SIGTERM, SIGINT, SIGKILL, SIGHUP, SIGPIPE)

>>> PLATFORM_PATH -- File path and filesystem differences:
  std::path::MAIN_SEPARATOR
  Path handling with \ vs / differences
  Case sensitivity in path comparisons
  Home directory resolution (~/ vs %USERPROFILE%)
  /dev/null vs NUL, /dev/tty vs CON

>>> PLATFORM_ENCODING -- Encoding/locale differences:
  OsStr/OsString conversion differences (WTF-8 vs UTF-16)
  Line ending handling (\n vs \r\n)
  Console/terminal encoding differences

>>> OTHER_MIGRATION_FEATURE -- Other migration/compatibility features:
  Cross-platform compatibility shims or fallback implementations
  Cargo/build-script handling for target-specific dependencies
  Normalizing permissions, process, terminal, or filesystem behavior across OSes
  Test helpers or adapters that validate equivalent behavior across platforms

>>> FALSE_POSITIVE -- Not platform-specific:
  Test infrastructure noise (test setup/teardown timing differences)
  Feature flags unrelated to platform (#[cfg(feature = ...)])
  Code coverage gaps (lines not reached due to test selection, not platform)
  Library version differences unrelated to OS
\end{lstlisting}
\end{tcolorbox}

\subsection{Masking the Target Implementation}

The masking procedure labels verified target-specific lines with \texttt{<linux\_specific>} for Windows-to-Linux tasks or \texttt{<windows\_specific>} for Linux-to-Windows tasks.

\begin{tcolorbox}[colback=blue!3, colframe=blue!50, title=Masking Prompt, breakable, fontupper=\scriptsize]
\begin{lstlisting}[style=patchbox]
You are provided with a file content where {marker} platform-specific lines have been labeled. You need to delete all {marker} platform-specific lines to create an incomplete code repository to test other software engineers' ability to re-implement it. Each of your str_replace tool call should satisfy ALL of the 3 requirements below:
    (1) Delete lines with prefix <{marker}> in some continuous code blocks.
    (2) Substitute the deleted lines and the contents surrounding the deleted lines with place holders to make the line deletion unobvious and make the remaining file content still look natural. This is to test other engineers' ability to find the hidden locations that have {marker}-specific problems from scratch. You must not substitute the deleted lines with `raise error` or `throw excpetion`, because `raise error / throw exception` is very obvious and easy to be found, which makes the created SWE task too trivial. 
    (3) Do not re-implement the deleted locations, as these locations are deleted intentionally to test other engineers' ability to find and re-implement them. Ensure each substituted code block (your generated place holder) will still cause problems or errors in subsequent code blocks during execution due to the specified deleted lines. Only when a deletion/substitution can cause error at subsequent code blocks or can make some regression tests fail will the edit be considered valid. Do not fix bugs anywhere else, your task is only to delete blocks with <{marker}>.
we will ask you to remove labeled blocks in a loop until there is no line with <{marker}>.
code file content:
```{language}
{cur_content}
```
Now do str_replace.
\end{lstlisting}
\end{tcolorbox}

\subsection{Migration Task Instruction}

\begin{tcolorbox}[colback=blue!3, colframe=blue!50, title=Migration Task Prompt, breakable, fontupper=\scriptsize]
\begin{lstlisting}[style=patchbox]
You are given a repository at /testbed. The repository was first developed on {source} but now you need to migrate it to {target}. Fix all platform compatibility issues on {target}. Your final goal is to pass all regression tests on {target}.
\end{lstlisting}
\end{tcolorbox}

\subsection{SWE-agent}

\begin{tcolorbox}[colback=blue!3, colframe=blue!50, title=System Prompt, breakable, fontupper=\scriptsize]
\begin{lstlisting}[style=patchbox]
You are a helpful assistant that can interact with a computer to solve tasks.
\end{lstlisting}
\end{tcolorbox}

\begin{tcolorbox}[colback=blue!3, colframe=blue!50, title=Instance Prompt, breakable, fontupper=\scriptsize]
\begin{lstlisting}[style=patchbox]
<uploaded_files>
{{working_dir}}
</uploaded_files>
I've uploaded a code repository in the directory {{working_dir}}. Consider the following PR description:

<pr_description>
{{problem_statement}}
</pr_description>

Can you help me implement the necessary changes to the repository so that the requirements specified in the <pr_description> are met?
I've already taken care of all changes to any of the test files described in the <pr_description>. This means you DON'T have to modify the testing logic or any of the tests in any way!
Your task is to make the minimal changes to non-tests files in the {{working_dir}} directory to ensure the <pr_description> is satisfied.
Follow these steps to resolve the issue:
1. As a first step, it might be a good idea to find and read code relevant to the <pr_description>
2. Figure out how to run regression tests in the repository.
3. Edit the source code of the repo to resolve the issue
4. Rerun the regression tests to confirm that the error is fixed!
5. Think about edgecases and make sure your fix handles them as well
Your thinking should be thorough and so it's fine if it's very long.

IMPORTANT NOTES:
1. Do not commit your edits, we will do it later. When you want to submit, just use the submit tool.
2. Do not use any git command. Git command is forbidden. If you use any git command your solution would be immediately judged as failed!
\end{lstlisting}
\end{tcolorbox}

\begin{tcolorbox}[colback=blue!3, colframe=blue!50, title=Submission Review Prompt, breakable, fontupper=\scriptsize]
\begin{lstlisting}[style=patchbox]
Thank you for your work on this issue. Please carefully follow the steps below to help review your changes.

1. If you made any changes to your code after running the reproduction script, please run the reproduction script again.
  If the reproduction script is failing, please revisit your changes and make sure they are correct.
  If you have already removed your reproduction script, please ignore this step.
2. Remove your reproduction script (if you haven't done so already).
3. If you have modified any TEST files, please revert them to the state they had before you started fixing the issue.
  You can do this with `git restore /path/to/test/file.py` (this is the only place you can use git; do not use git anywhere else, otherwise your solution would be marked as failed). Use below <diff> to find the files you need to revert.
4. Run the submit command again to confirm.

Here is a list of all of your changes:

<diff>
{{diff}}
</diff>
\end{lstlisting}
\end{tcolorbox}

\subsection{Claude Code}

\begin{tcolorbox}[colback=blue!3, colframe=blue!50, title=Claude Code, breakable, fontupper=\scriptsize]
\begin{lstlisting}[style=patchbox]
You are given a code repository in the current directory (pwd && ls ./).
The issue description is:
{description}
=================================================
Your task is to fix the issue with the following steps:
(1) explore the source codes to find relevant files.
(2) figure out how to run regression tests.
(3) edit the source codes to fix the bug.
(4) rerun regression tests to see whether all linux-compatibility issues are fixed. If not, go back to explore and fix.
(5) remember to delete the test cases you write at last; if you modified the regression test files, git restore it. Note this is the only place you can use git command; if you use git command elsewhere your solution will be immediately judged as failed!
=================================================
Notes:
(1) Please do not commit your edits. We will do it later.
(2) Do not use any git command except git restore the test files. Git command other than git restore the test files is forbidden. If you use any git command your solution would be immediately judged as failed!
(3) In our setting, each of your response must contain exactly one text content and one tool call. If you want to submit your result, just output one reponse with only text content and without tool call -- a response without tool call means stop and submission. If you have not finished, each of your response must have one tool call to make the session continue.
\end{lstlisting}
\end{tcolorbox}

\subsection{OpenHands}

\begin{tcolorbox}[colback=blue!3, colframe=blue!50, title=OpenHands, breakable, fontupper=\scriptsize]
\begin{lstlisting}[style=patchbox]
I have access to a code repository in the directory {{ instance.repo_path }} . You can explore and modify files using the available tools. Consider the following issue description:

<issue_description>
{{ instance.problem_statement }}
</issue_description>

Can you help me implement the necessary changes to the repository so that the requirements specified in the <issue_description> are met?
I've already taken care of all changes to any of the test files described in the <issue_description>. This means you DON'T have to modify the testing logic or any of the tests in any way!
Also the development environment is already set up for you (i.e., all dependencies already installed), so you don't need to install other packages.
Your task is to make the minimal changes to non-test files in the {{ instance.repo_path }} directory to ensure the <issue_description> is satisfied.

Follow these phases to resolve the issue:

Phase 1. READING: read the problem and reword it in clearer terms
   1.1 If there are code or config snippets. Express in words any best practices or conventions in them.
   1.2 Hightlight message errors, method names, variables, file names, stack traces, and technical details.
   1.3 Explain the problem in clear terms.
   1.4 Enumerate the steps to reproduce the problem.
   1.5 Hightlight any best practices to take into account when testing and fixing the issue

Phase 2. RUNNING: install and run the regression tests on the repository
   2.1 Follow the readme
   2.2 Iterate and figure out how to run the regression tests

Phase 3. EXPLORATION: find the files that are related to the problem and possible solutions
   3.1 Use `grep` to search for relevant methods, classes, keywords and error messages.
   3.2 Identify all files related to the problem statement.
   3.3 Propose the methods and files to fix the issue and explain why.
   3.4 From the possible file locations, select the most likely location to fix the issue.

Phase 4. FIX ANALYSIS: state clearly the problem and how to fix it
   4.1 State clearly what the problem is.
   4.2 State clearly where the problem is located.
   4.3 State clearly how the test reproduces the issue.
   4.4 State clearly the best practices to take into account in the fix.
   4.5 State clearly how to fix the problem.

Phase 5. FIX IMPLEMENTATION: Edit the source code to implement your chosen solution.
   5.1 Make minimal, focused changes to fix the issue.

Phase 6. VERIFICATION: Test your implementation thoroughly.
   6.1 Run your reproduction script to verify the fix works.
   6.2 Add edge cases to your test script to ensure comprehensive coverage.
   6.3 Run existing tests related to the modified code to ensure you haven't broken anything.

7. FINAL REVIEW: Carefully re-read the problem description and compare your changes with the base commit {{ instance.base_commit }}.
   7.1 Ensure you've fully addressed all requirements.
   7.2 Run any tests in the repository related to:
     7.2.1 The issue you are fixing
     7.2.2 The files you modified
     7.2.3 The functions you changed
   7.3 If any tests fail, revise your implementation until all tests pass

NOTE: 
 - Be thorough in your exploration, testing, and reasoning. It's fine if your thinking process is lengthy - quality and completeness are more important than brevity.
 - If you have modified test files, git restore them before submission. (Note this is the only place where you can use git command. If you use git command elsewhere your solution will be immediately judged as failed.)
 - You don't need to git commit. If you want to submit just use the `finish` tool.
 - Do not use any git command except git restore the test files. Git command is forbidden. If you use any git command your solution would be immediately judged as failed!
\end{lstlisting}
\end{tcolorbox}

\subsection{Win-agent}

\begin{tcolorbox}[colback=blue!3, colframe=blue!50, title=Win-agent, breakable, fontupper=\scriptsize]
\begin{lstlisting}[style=patchbox]
You are given a code repository at C:\testbed
You are in Windows powershell
You need to solve the following problem:
_PROBLEM_STATEMENT_
=================================================
Follow these steps to solve the problem:
1) Explore the repository to find relevant codes.
2) Figure out how to run regression tests in the repository.
3) Edit codes to solve the problem.
4) Run regression tests to validate your fix. If the test fails, go back to explore and edit again.
5) If all the tests can pass, remember to delete any test files you create and git restore the test files if you modified any. Then use submit tool call to submit. Note git restore the test files is the only git command you can use. Do not use git command anywhere else, otherwise your solution would be immediately judged as failed.
=================================================
Note:
1) Each of your response should contain exactly one tool call.
2) Do not git commit, if you want to submit just use the submit tool.
3) Do not use git command anywhere except git restore the test files. Git command is forbidden. If you use git command, your solution would be immediately judged as failed.
4) As you are in a powershell terminal, DO NOT use `exit $LASTEXITCODE` in your command as it will terminate this conversation.
\end{lstlisting}
\end{tcolorbox}

\end{document}